\documentclass[
  prx,
  twocolumn,
  superscriptaddress,
  showpacs]
  {revtex4-2} %{revtex4-2}

  \usepackage{graphicx}
  \usepackage{amsmath,amssymb}
  \usepackage{dcolumn}% Align table columns on decimal point
  \usepackage{multirow}

\newcommand{\kB}{k_{\text{B}}}

\newcommand{\pMI}{{\em pMI~}}

\begin{document}

\title{Formation enthalpies of Al-Mn-Pd and the structure of the $i$-AlMnPd quasicrystal}

\author{Marek Mihalkovi\v c}
\affiliation{Institute of Physics, Slovak Academy of Sciences, 84511 Bratislava, Slovakia}
\email{mihalkovic@savba.sk}
\author{Michael Widom}
\affiliation{Department of Physics, Carnegie Mellon University, Pittsburgh, PA  15213, USA}
\email{widom@cmu.edu}
\date{\today}

\begin{abstract}
This paper reports formation enthalpies of phases in the Al-Mn-Pd ternary alloy system as calculated from first principles using electronic density functional theory. We consider all crystal structures as reported in the assessed phase diagrams of the ternary and its binary alloy subsystems (Al-Mn, Al-Pd, and Mn-Pd), as well as additional reported or hypothetical structures. Icosahedral and decagonal quasicrystalline approximants are among the structures that we predict to be stable, or nearly so. Our results suggest the need for careful experimental reexamination of phase stability in each of the alloy systems, in tandem with further efforts to refine crystallographic and {\em ab-initio} structures.
\end{abstract}


\maketitle

\section{Introduction}

AlMnPd is one of the most studied quasicrystal forming systems. It has an icosahedral phase located around composition Al$_{70}$Mn$_{10}$Pd$_{20}$ and a decagonal phase close to Al$_{70}$Mn$_{20}$Pd$_{10}$~\cite{Grushko1999}. The icosahedral phase lies close in composition to the first-discovered quasicrystal~\cite{Shechtman1984}, Al$_{86}$Mn$_{14}$, and it can be prepared by equilibrium methods~\cite{Boudard}. Despite that, to-date no existing structure model passed the scrutiny of ab--initio evaluation of its stability against competing crystal phases, thus motivating the present study.

We match the assessed experimentally observed phases as reported by the American Society of Metals~\cite{ASM} with crystallographic structures as reported by the Inorganic Crystal Structure Database~\cite{ICSD}. Our goal is to identify crystal structures for each reported phase, and to validate this identification through first-principles total energy calculations. Structures are identified through a combination of their Pearson names, space group and other identifiers. Phases are identified through their names, compositions, and other pertinent information. Icosahedral and decagonal quasicrystals are represented by ``approximant'' structures that spontaneously form during simulations.

Our calculated formation enthalpies reveal numerous conflicts withreports of structure and phase stability. The origins and potential resolutions of the conflicts are discussed. Remarkably, the only ternary structures predicted to be stable at low temperatures are large unit cell quasicrystalline approximants.

\section{Methods}
\label{sec:methods}

Our calculations follow methods outlined in a prior paper~\cite{Mihal04}.  We utilize VASP~\cite{Kresse96} to carry out first principles density functional theory (DFT) total energy calculations in the PBE generalized gradient approximation~\cite{PBE}. We adopt projector augmented wave potentials~\cite{Blochl94,Kresse99} and maintain a fixed energy cutoff of 270 eV (the default for Mn).  All calculations involving Mn considered the possibility of spin polarization. We relax all atomic positions and lattice parameters using the Normal precision setting, and increase our $k$-point densities until energies have converged to within 1 meV/atom, then carry out a final static calculation using the tetrahedron integration method. 

Given total energies for a variety of structures, we calculate the enthalpy of formation $\Delta H_{\rm for}$ which is the enthalpy of the structure relative to a tie-line connecting the ground state configurations of the pure elements~\cite{Mihal04}.  Formally, for a compound of stoichiometry Al$_x$Mn$_y$Pd$_z$ where $x+y+z=1$, we define 
\begin{equation}
\label{eq:H}
\Delta H_{\rm for}=H({\rm Al}_x{\rm Mn}_y{\rm Pd}_z)-(x H({\rm Al})+ y H({\rm Mn}) + z H({\rm Pd}))
\end{equation}
where all enthalpies are per atom.  Vertices of the convex hull of $\Delta H_{\rm for}$ constitute the predicted low temperature stable structures.  For structures that lie above the convex hull, we calculate the instability energy $\Delta E$ as the enthalpy relative to the convex hull.

\begin{figure}
  \includegraphics[width=0.45\textwidth]{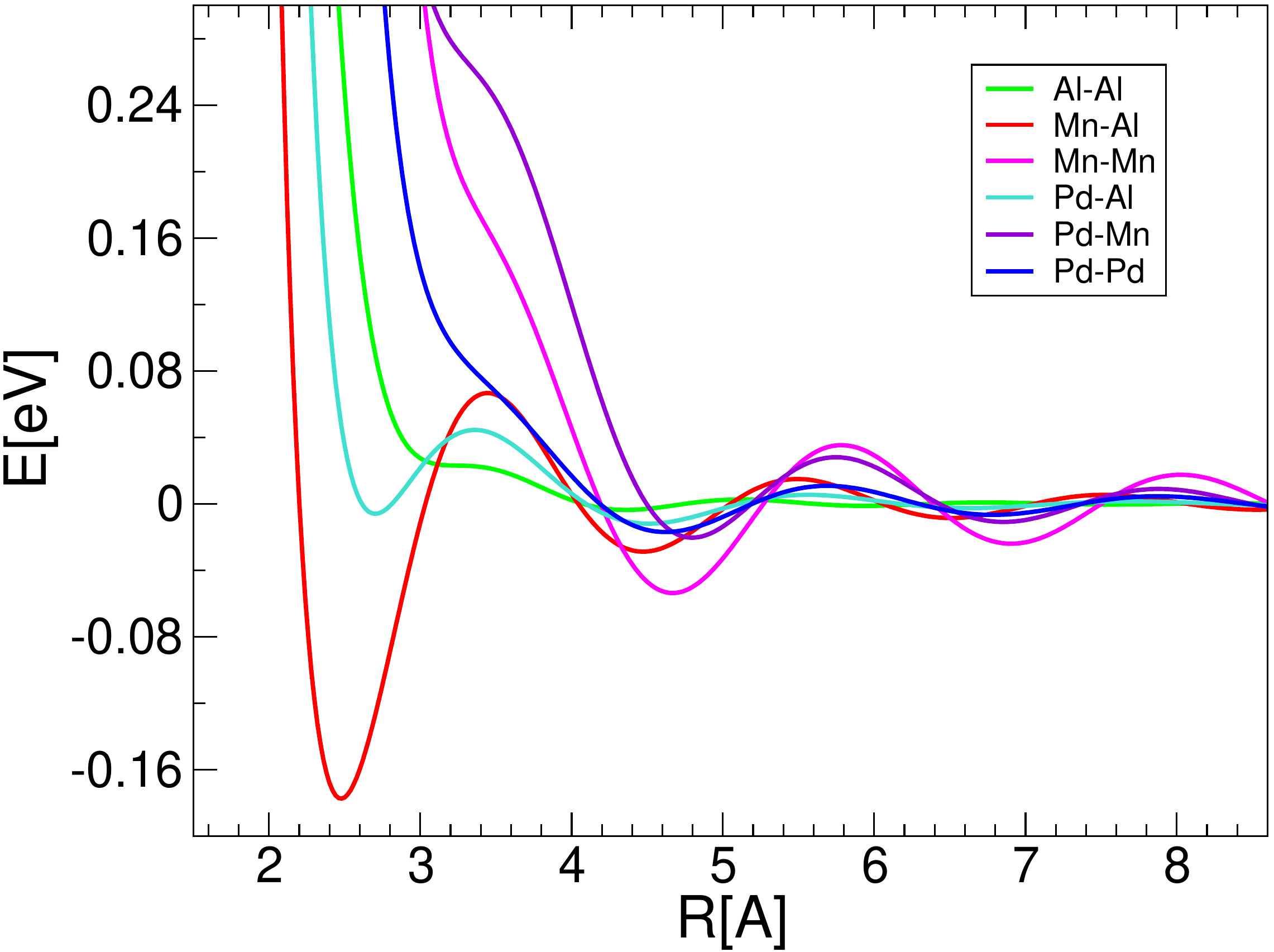}
  \caption{\label{fig:pp} Pair potentials as given by Eq.~(\ref{eq:oscil6}) with fitted parameters as in Table~\ref{tab:eopp}.}
\end{figure}

Our structures are drawn from the ASM phase diagram database ~\cite{ASM} and from the Inorganic Crystal Structure Database~\cite{ICSD} (ICSD), supplemented with original publications. When presented with mutiple structure possibilities, or mixed site occupancy, we examine plausible structures and report the most energetically favorable. We sometimes refine structures using Monte Carlo annealing and relaxation subject to empirically oscillating pair potentials~\cite{EOPP}. Our potentials are fitted to the form
\begin{equation}
  \label{eq:oscil6}
  V(r) = \frac{C_1}{r^{\eta_1}} + \frac{C_2} {r^{\eta_2}} \cos(k_* r + \phi_*)
\end{equation}
with parameters as given in Table~\ref{tab:eopp} and Fig.~\ref{fig:pp}. We also perform Monte Carlo/molecular dynamics simulations supplemented with replica exchange~\cite{i-AlCuFe} to optimize quasicrystalline structures.

\begin{table}
  \begin{tabular}{c|cccccc}
    \hline
    & $C_1$ & $\eta_1$ & $C_2$ & $\eta_2$ & $k_*$ & $\phi_*$ \\
    \hline
Al--Al  &  2497.0             & 9.9797 &  -2.1972 & 4.1609 & 3.9470 & 1.5270 \\
Al--Mn  & 35935               &15.683  &   3.9725 & 3.2700 & 3.1220 & 1.5137 \\
Al--Pd  &  2052.7             &10.211  &   5.9551 & 4.0744 & 3.0032 & 1.8990 \\
Mn--Mn  &1.3871$\times$10$^5$ &11.753  &  -1.5400 & 2.1493 &-2.8182 & 0.7226 \\
Mn--Pd  &   606.71            & 6.5741 &  -6.2022 & 3.2063 &-3.0004 & 1.8224 \\
Pd--Pd  & 16446               &10.572  &  -1.0907 & 2.6558 &-2.9027 & 0.9748 \\
  \end{tabular}
  \caption{\label{tab:eopp} Fitted parameters for Al--Mn--Pd EOPP potentials.}
\end{table}

One of our goals is to screen the experimentally assessed phase diagrams for conflicts with DFT evaluations of stability. We flag cases where the assessment claims low temperature (LT) or room temperature (RT) stability but DFT indicates an enthalpy lying above the convex hull. Since we neglect entropic effects, $\Delta E$ values of order $\kB T\approx 26$ meV/atom could conceivably be overcome, but more likely the reported crystallographic structure needs refinement, or more careful equilibration is required.

\section{Al-Mn-Pd Ternary Alloy System}

To evaluate stability of a ternary alloy system we need to examine all three of its binary subsystems, as well as all three constituent elements.

\subsection{Pure elements}

Elemental Al and Pd each take the face centered cubic structure (FCC, Pearson type cF4, space group Fm$\bar{3}$m) at all temperatures up to their melting points. Mn, in contrast, takes one of the most complex elemental structures at low temperature, $\alpha$-Mn (Pearson type cI58, space group I$\bar{4}$3m), then passes through high temperature (HT) phases $\beta$-Mn (Pearson type cP20, space group P4$_1$32) and $\gamma$-Mn (FCC) until reaching body centered cubic $\delta$-Mn (BCC, Pearson type cI2, space group Im$\bar{3}$m). Our calculations find that $\alpha$-Mn exhibits a type of antiferromagnetism, with moments of 2.6$\mu_B$ on the 2a site and -1.66$\mu_B$ on the 8c sites. $\beta$- and $\gamma$-Mn prove at most weakly magnetic. $\delta$-Mn favors ferromagnetism in its electronic ground state, although as a high temperature phase, it exists only above its hypothetical Curie point. In summary we find complete agreement between assessed and DFT-predicted stability of the three elements.

\subsection{Binary subsystems}

\subsubsection{Al-Mn}

\begin{figure}
  \includegraphics[width=0.49\textwidth]{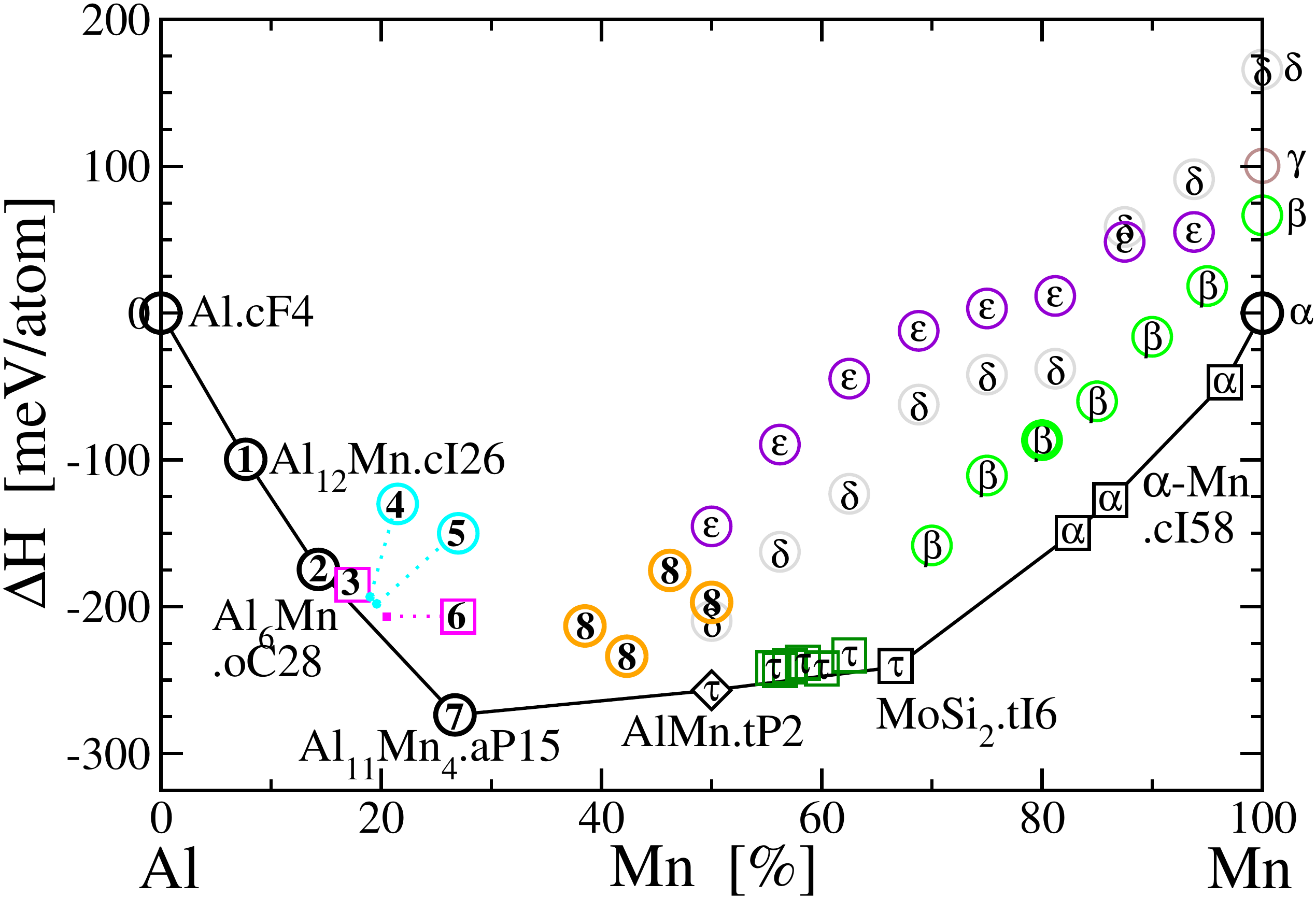}
  \caption{\label{fig:Al-Mn} Formation enthalpies of the Al-Mn alloy system. Plotting symbols are: heavy circle assessed RT/LT stable; light circle assessed HT stable; diamond assessed metastable; square hypothetical or unassessed structure. Stable structures on the convex hull are labeled. See Table~\ref{tab:Al-Mn} for details.}
\end{figure}

The ASM assessed phase diagram of Al$_{1-x}$Mn$_x$~\cite{ASM,ASM_Al-Mn} contains numerous line compounds (phases of precisely fixed composition) with complex structures in the range of $x_{\rm Mn}=0.08-0.28$. Our DFT-calculated enthalpies of formation (See Fig.~\ref{fig:Al-Mn} and Table~\ref{tab:Al-Mn}) validate the low temperature stable structures as Al$_{12}$Mn as Pearson type cI26, Al$_6$Mn as Pearson type oC28, and Al$_{11}$Mn$_4$ as Pearson type aP15.

Al$_{57}$Mn$_{12}$.cP138 is a metastable icosahedral quasicrystal approximant that can be stabilized through the addition of Si~\cite{Cooper,ElserHenley}. Its structure is a body-centered cubic packing of Mackay icosahedra with empty centers. The symmetry is reduced from BCC to primitive cubic by chemical ordering of the clusters' third shells. %Marek: check "chemical ordering".

The assessed diagram lists two phases~\cite{Kreiner, Shoemaker} $\lambda$-Al$_{230.8}$Mn$_{53.3}$.hP586 rt (space group P6$_3$/m) and $\mu$-Al$_{226.6}$Mn$_{55}$.hP563 (space group P6$_3$/mmc). They lie close in composition and remain stable over an extended temperature range, which is thermodynamically improbable~\cite{Okamoto1991}. Indeed, a more recent assessment~\cite{Grushko2008} continues the debate over the precise details of the phase diagram, with the main points of agreement that $\mu$ is more Mn-rich than $\lambda$ and extends to higher temperature. The two phases are related by $a_\lambda=\sqrt{2}a_\mu$ and $c_\lambda=c_\mu/2$. All atoms in $\lambda$ and $\mu$ belong to Mn-centered 13-atom icosahedral clusters, with the exception of two Al$_9$Mn clusters in which the Mn atoms are surrounded by tri-capped trigonal prisms. The icosahedra have various orientations, so we do not consider these phases as quasicrystal approximants. According to our calculations, neither is stable in the limit of low temperature. Both structures exhibit mixed or partial occupancy and hence should be regarded as high temperature phases. At slightly greater Mn concentration we have a decagonal quasicrystal approximant, Y-Al$_3$Mn.oP156~\cite{ShiLiMaKuo}. Curiously, we find the energetically best structure of the Y-phase occurs at a composition close to Al$_4$Mn.

\begin{table}
  \caption{\label{tab:Al-Mn} Selected enthalpies of the Al-Mn binary system. Labels specify plotting symbols on Fig.~\ref{fig:Al-Mn}. Phase names are taken from ASM~\cite{ASM} or prototype. Space groups refer to observed symmetries and may differ from DFT realizations due to mixed or partian occupation. ASM stabilities are listed as RT if they extend to lowest reported ASM temperature, HT are high temperature only, and MS is assessed metastable. DFT energies (meV/atom) are formatio enthalpy $\Delta H$ and stability relative to competing phases $\Delta E$. Bolded stability and $\Delta E$ indicate conflicts between DFT and assessed stability. Bolded phase names indicate a DFT predicted stable structures.}
  \begin{tabular}{l|rrr|r|rrr}
    Label & Phase   & Pearson & Space & ASM       & \multicolumn{3}{c}{DFT} \\
          & name    & symbol  & group & stability & x$_{\rm Mn}$ & $\Delta H$ & $\Delta E$ \\
    \hline
   1  & Al$_{12}$Mn & cI26    & Im$\bar{3}$   & RT &  7.7 & -98 &  -6 \\
   2  & Al$_6$Mn    & oC28    & Cmcm          & RT & 14.3 & -171 & -13 \\
   3  & Al$_{57}$Mn$_{12}$ & cP138 & Pm$\bar{3}$  & NA & 17.4 & -182 &  14 \\
   4  & $\lambda$-Al$_4$Mn & hP568 & P6$_3$/m & {\bf RT} & 19.0 & -192 & {\bf 17} \\
   5  & $\mu$-Al$_4$Mn & hP563   & P6$_3$/mmc     & {\bf RT} & 19.6 & -198 & {\bf 17} \\
   6  & Y-Al$_3$Mn    & oP156   & Pna2$_1$      & NA & 20.5 & -204 &  18 \\
%   6  & Al$_3$Fe    & mC102   & C2/m          & NA & 23.5 & -239 &  14 \\
   7  & $\nu$-Al$_{11}$Mn$_4$ & aP15 & P$\bar{1}$   & RT & 26.7 & -272 & -33 \\
   8  & Al$_8$Mn$_5$ & hR26   & R3m           & {\bf RT} & 50.0 & -194 &  {\bf 62} \\
 $\tau$ & AlMn & tP2     & P4/mmm      & {\bf MS} & 50.0 & -256 & {\bf -2} \\
 $\tau$ & {\bf MoSi$_2$} & tI6  & I4/mmm & {\bf NA} & 66.7 & -240 & {\bf -24} \\
   25 & $\epsilon$-Al$_{55}$Mn$_{45}$ & hP2 & P6$_3$/mmc & HT & 50.0 & -144 & 112 \\
   17 & $\delta$-Al$_{45}$Mn$_{55}$ & cI2 & Im$\bar{3}$m & HT & 50.0 & -211 & 45 \\
   11 & $\beta$-Mn  & cP20    & P4$_1$32      & {\bf RT} & 80.0 & -86 & {\bf 79} \\
   14 & $\alpha$-Mn & cI58    & I$\bar{4}$3m  & {\bf NA} & 82.8 & -150 & {\bf -2} \\ 
  \end{tabular}
\end{table}

Several solid solutions with extended composition ranges exist beyond $x_{\rm Mn}=0.3$. The assessed Al15Mn11.hR26, also known as Al8Mn5.hR26~\cite{Ellner}, is a high temperature phase up to $\sim$50\% Mn at which point it is assessed as stable down to low temperatures. DFT shows it could be HT stable but not LT, at least not with the site occupancy as indicated by experiment~\cite{Thimmaiah}.

We find instead that a tP2 structure, assessed to be metastable, seems to be the true LT structure, with ferromagnetic order and $c/a>1$. We also predict LT stability of an AlMn$_2$.tI6 phase based on the prototype MoSi$_2$. This prototype is prevalent among Al-rich intermetallics~\cite{tI6} but has not been reported experimetally in Al-Mn, likely due to the metastability of the high temperature solid solution phases. The tP2 structure evolves continuously into tP6 through a sequence of layered structures alternating Al and Mn planes that we collectively label $\tau$. Our DFT calculations show that the endpoints are LT stable, while the intermediate structures lie within 12 meV or less of the convex hull. AlMn.tP2 has magnetic moments of 2.3$\mu_{\rm B}$/Mn atom, while AlMn$_2$.tI6 is predicted to be nonmagnetic.

The highest temperature Mn phase, $\delta$-Mn.cI2 extends in the assessed diagram from 40-100\% Mn, but is interrupted by an hP2 structure in the range 55-75\% Mn. Meanwhile, the HT $\beta$-Mn.cP20 phase extends at intermediate temperatures from 60-100\% Mn, but becomes RT stable around 80\% Mn. However, cP20 lies significantly above the convex hull over the entire composition range. Rather, the $\alpha$-Mn.cI58 phase meets the convex hull if we substitute Al on the 2a site, the 8c site, or both. Presumably the $\beta$-Mn.cP20 variant is entropically stabilized at high temperatures and remains metastable down to low temperatures, preventing the observation of the more stable $\alpha$ variant.

We examined stability of the hP2 phase (not shown) and found that it lies strictly above cI2 for 50-85\%Mn but falls below it in the limit of high Mn content. This behavior does not reproduce the experimental observation and likely shows the need for more rigorous approach to the {\em ab-initio} study of high temperature phases through inclusion of vibrational entropy, chemical substitution, or vacancies~\cite{Wolverton2001,i-AlCuFe}.

\subsubsection{Al-Pd}

The Al-rich region of the assessed diagram~\cite{ASM,ASM_Al-Pd_Li2006} reports a hexagonal RT stable phase, Al$_4$Pd, of known lattice parameters but unknown structure. DFT finds the prototype Al$_4$Pt.hP102 is predicted nearly stable ($\Delta E=9$ meV/atom) in a realization with 90 atoms, and it matches the experimental lattice parameters. The vacancies occur in mutually exclusive pairs of Al sites confined within channels, similar to those found in other transition metal aluminides~\cite{Al5Fe2}, and these can provide a source entropy that might stabilize the structure at elevated temperatures.

With increasing Pd concentration, the assessed diagram lists an orthorhombic HT stable Al$_3$Pd phase with known lattice constants but unknown structure. We checked a reported 16~\AA~ periodic decagonal structure~\cite{Hiraga-Al-Pd}, $\xi'$, with similar lattice constants and found it to be unstable by just $\Delta E=23$ meV/atom. That is sufficiently close to the convex hull that it could be stabilized by the entropy of mixed and partial occupation.

We also tested the stability of the reported metastable crystal structure Al$_2$Pd.cF12 and find it to be unstable by $\Delta E=39$ meV/atom.  $\delta$-Al$_3$Pd$_2$.hP5 is LT according to both the assessed diagram and DFT. The assessed diagram lists two Al$_{21}$Pd$_8$ phases, one is orthorhombic and is HT stable but no structure is given; the other is LT stable with structure tI116. DFT confirms LT stability of tI116.

\begin{figure}
  \includegraphics[width=0.45\textwidth]{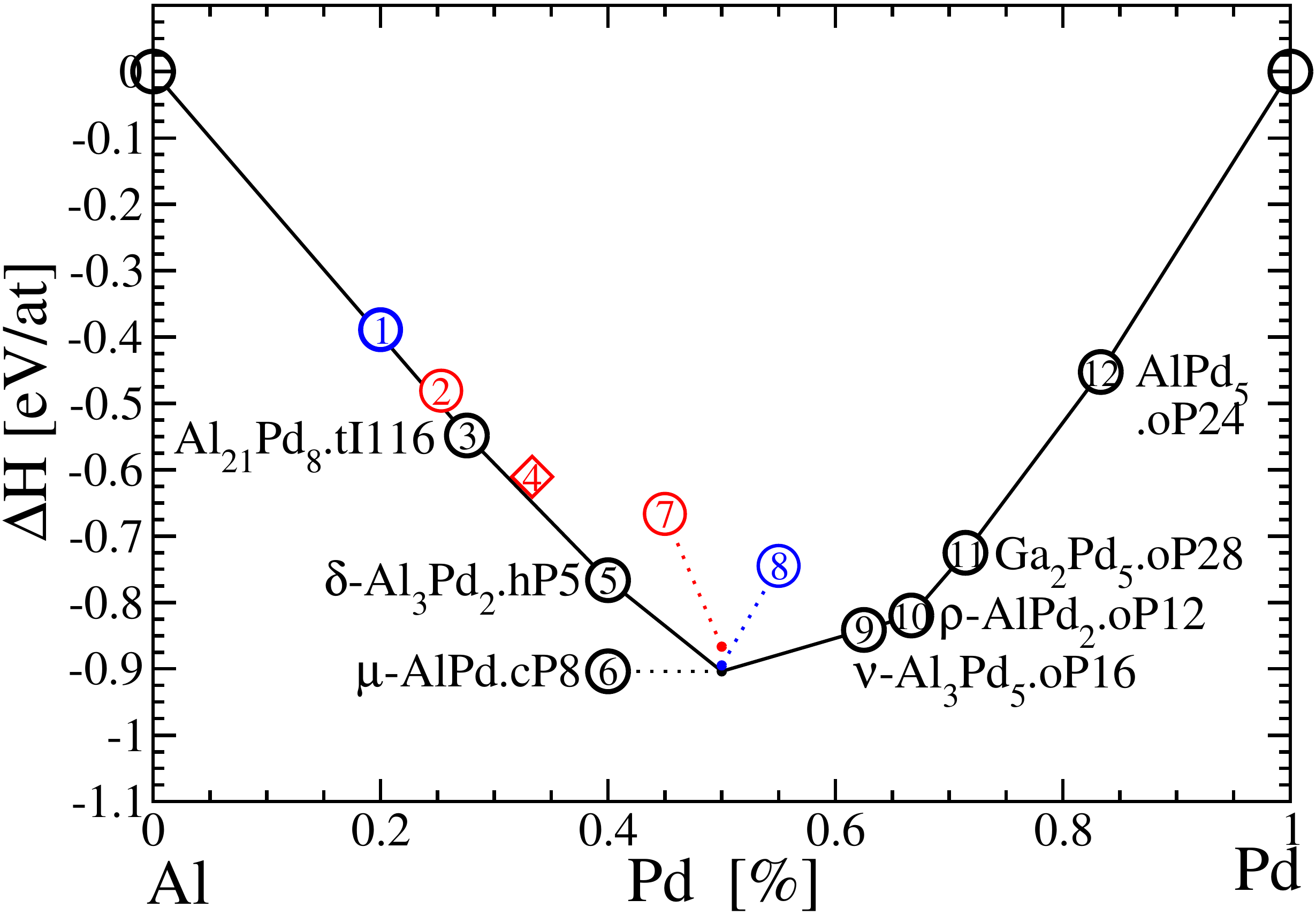}
  \caption{\label{fig:Al-Pd} Formation enthalpies of the Al-Pd alloy system. Plotting symbols as in Fig.~\ref{fig:Al-Mn}}
\end{figure}

\begin{table}
  \caption{\label{tab:Al-Pd} Selected enthalpies of the Al-Pd binary system. Labels specify plotting symbols on Fig.~\ref{fig:Al-Pd}. Other details as in Table~\ref{tab:Al-Mn}.}
  \begin{tabular}{l|rrr|r|rrr}
    Label & Phase   & Pearson & Space & ASM       & \multicolumn{3}{c}{DFT} \\
          & name    & symbol  & group & stability & x$_{\rm Pd}$ & $\Delta H$ & $\Delta E$ \\
   \hline
   1  & $\lambda$-Al$_4$Pd  & hP102 & P3c1        & {\bf RT} & 25.0 & -389 & {\bf 9} \\
   2  & $\xi'$-Al$_3$Pd & oP300   & Pna2$_1$& HT & 25.3 & -481 &   23 \\
   3  & Al$_{21}$Pd$_8$ & tI116 & I4$_1$/a  & RT & 27.6 & -549 &  -16 \\
   4  & Al$_2$Pd     & cF12  & Fm$\bar{3}$m & MS & 33.3 & -611 &   39 \\
   5  & $\delta$-Al$_3$Pd$_2$ & hP5 & P$\bar{3}$m1 & RT & 40.0 & -766 & -21 \\
   6  & $\mu$-AlPd   & cP8   & P2$_1$3      & RT & 50.0 & -904 &   -9 \\
   7  & AlPd         & cP2   & Pm$\bar{3}$m & HT & 50.0 & -867 &   37 \\
   8  & $\beta$-AlPd & hR26  & R$\bar{3}$   & {\bf RT} & 50.0 & -895 &{\bf 9} \\
   9  & $\nu$-Al$_3$Pd$_5$ & oP16 & Pbam    & HT & 62.5 & -841 &   -1 \\
   10 & $\rho$-AlPd$_2$ & oP12 & Pnma       & RT & 66.7 & -820 &  -32 \\
   11 & {\bf Ga$_2$Pd$_5$} & oP28  & Pnma         & RT & 71.4 & -725 &   -10 \\
   10 & AlPd$_5$     & oP24  & Pnma         & RT & 83.3 & -453 &  -30 \\
  \end{tabular}
\end{table}

One phase diagram~\cite{ASM_Al-Pd_Li2006} lists a single HT phase from 45-55\% Pd, while another~\cite{ASM_Al-Pd_Okamoto2003} lists the same HT phase, plus two RT phases that are listed without offering structures. Other references~\cite{Matkovic,Ferro} list AlPd.cP8 (Al-rich) and AlPd.hR26. DFT finds stoichiometric AlPd.cP8 to be stable and AlPd.cP2 and $\beta$-AlPd.hR26 to be potentially HT. Curiously, all three AlPd phases are canonical cell tilings (CCT~\cite{clhcct}) in which Al atoms occupy even nodes, and Mn occupy odd nodes. $\beta$- and $\mu$- are ABC tilings, while the cP2 structure is a pure A tiling.

At 62-72\% Mn, the assessed diagram claims $\nu$-Al$_3$Pd$_5$.oP16 is HT but DFT predicts it to be LT stable. DFT confirms stability of $\rho$-AlPd$_2$.oP12. The RT Al$_2$Pd$_5$ phase has no assessed structure, but DFT predicts prototype Ga$_2$Pd$_5$.oP28 to be stable. Finally, Pd.cF4 exists as a solid solution from 80-100\% Pd.
  
\subsubsection{Mn-Pd}

The Mn-Pd phase diagram~\cite{ASM,ASM_Mn-Pd} is dominated by four phases with broad composition ranges. Generically, such phases should either exist only at high temperatures, or else they should narrow to a single specific composition at low temperature. This is because either chemical substitution or vacancies are required to create a composition range, and these would create entropy in the limit $T\to 0$K, violating the third law of thermodynamics~\cite{Abriata04,Okamoto1991}.

\begin{figure}
  \includegraphics[width=0.45\textwidth]{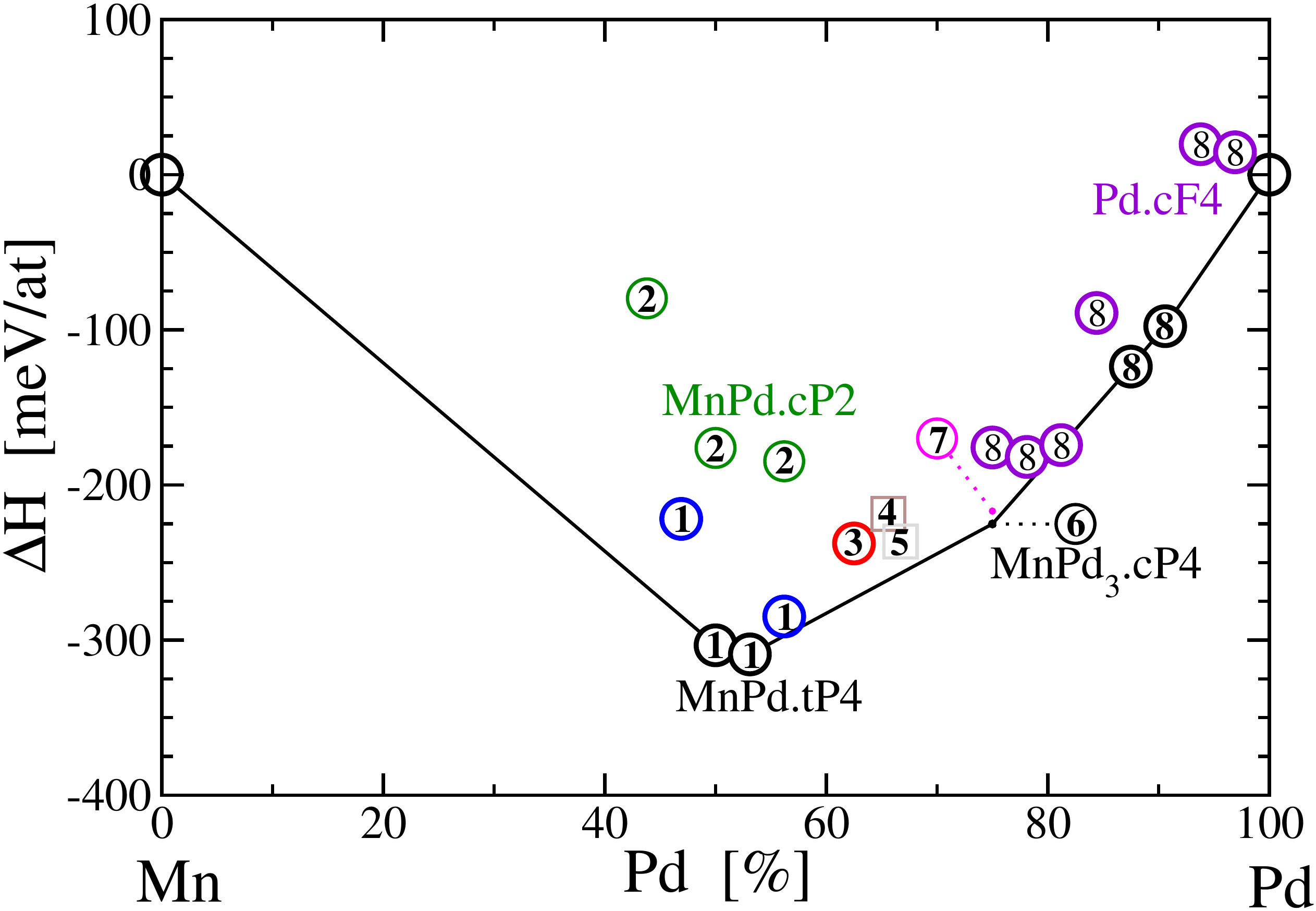}
  \caption{\label{fig:Mn-Pd} Formation enthalpies of the Mn-Pd alloy system. Plotting symbols as in Fig.~\ref{fig:Al-Mn}}
\end{figure}

\begin{table}
  \caption{\label{tab:Mn-Pd} Selected enthalpies of the Mn-Pd binary system. Labels specify plotting symbols on Fig.~\ref{fig:Mn-Pd}. Other details as in Table~\ref{tab:Al-Mn}.}
  \begin{tabular}{l|rrr|r|rrr}
    Label & Phase   & Pearson & Space & ASM       & \multicolumn{3}{c}{DFT} \\
          & name    & symbol  & group & stability & x$_{\rm Pd}$ & $\Delta H$ & $\Delta E$ \\
   \hline
   1  & MnPd     & {\bf tP4} & P4/mmm       & RT & 50.0 & -303 &  -12 \\
   2  & MnPd         & cP2   & Pm$\bar{3}$m & HT & 50.0 & -176 &  127 \\
   3  & Mn$_3$Pd$_5$ & oC16  & Cmmm    & {\bf RT} & 62.5 & -238 & {\bf 35} \\
   4  & Mn$_{11}$Pd$_{21}$ & tP32 & P4/mmm  & ?? & 65.6 & -219 &   42 \\
   5  & MnPd$_2$     & oP12  & Pnma         & ?? & 66.7 & -237 &   21 \\
   6  & MnPd$_3$     & cP4   & Pm$\bar{3}$m & HT & 75.0 & -225 &   -8 \\
   7  & Al$_3$Zr     & tI16  & I4/mmm       & {\bf RT} & 75.0 & -217 & {\bf 8}\\
   8 & Pd           & cF4   & Fm$\bar{3}$m & {\bf RT}& 75.0 & -176 &{\bf 49} \\
  \end{tabular}
\end{table}

Around 50\% Pd the assessed diagram shows a low temperature phase MnPd.tP2. DFT verifies this structure to be stable but with with antiferromagnetic order creating a $\sqrt{2}\times\sqrt{2}$ cell, hence the Pearson type tP4. There is also a HT phase MnPd.cP2; DFT finds this phase to be unstable by $\Delta E = 145$ meV/atom suggesting possible metastability. The assessed diagram lists Mn$_3$Pd$_5$.oC16 as RT, but DFT predicts it to be unstable by $\Delta E=27$ meV/atom suggesting possible entropic stabilization at HT. ASM also lists a stable Mn-rich variant of Mn$_3$Pd$_5$ with unknown structure.

At 75\% Pd, ASM lists two different MnPd$_3$ phases as stable: one at the low Pd limit of an FCC solid solution, and also one with prototype Al$_3$Zr.tI16~\cite{Rodic}. We predict both to be unstable or HT. Finally, the assessed diagram shows the FCC solid solution as RT over the range 75-100\% Pd. DFT instead predicts a sequence of optimized tetragonal structures at greater Pd \% that lie on or near the convex hull. At the low Pd limit, MnPd$_3$, ASM lists cP4 at HT, but DFT predicts it to be stable. 

\subsection{Ternary Al-Mn-Pd}

\begin{figure}[h]
  \includegraphics[width=0.45\textwidth]{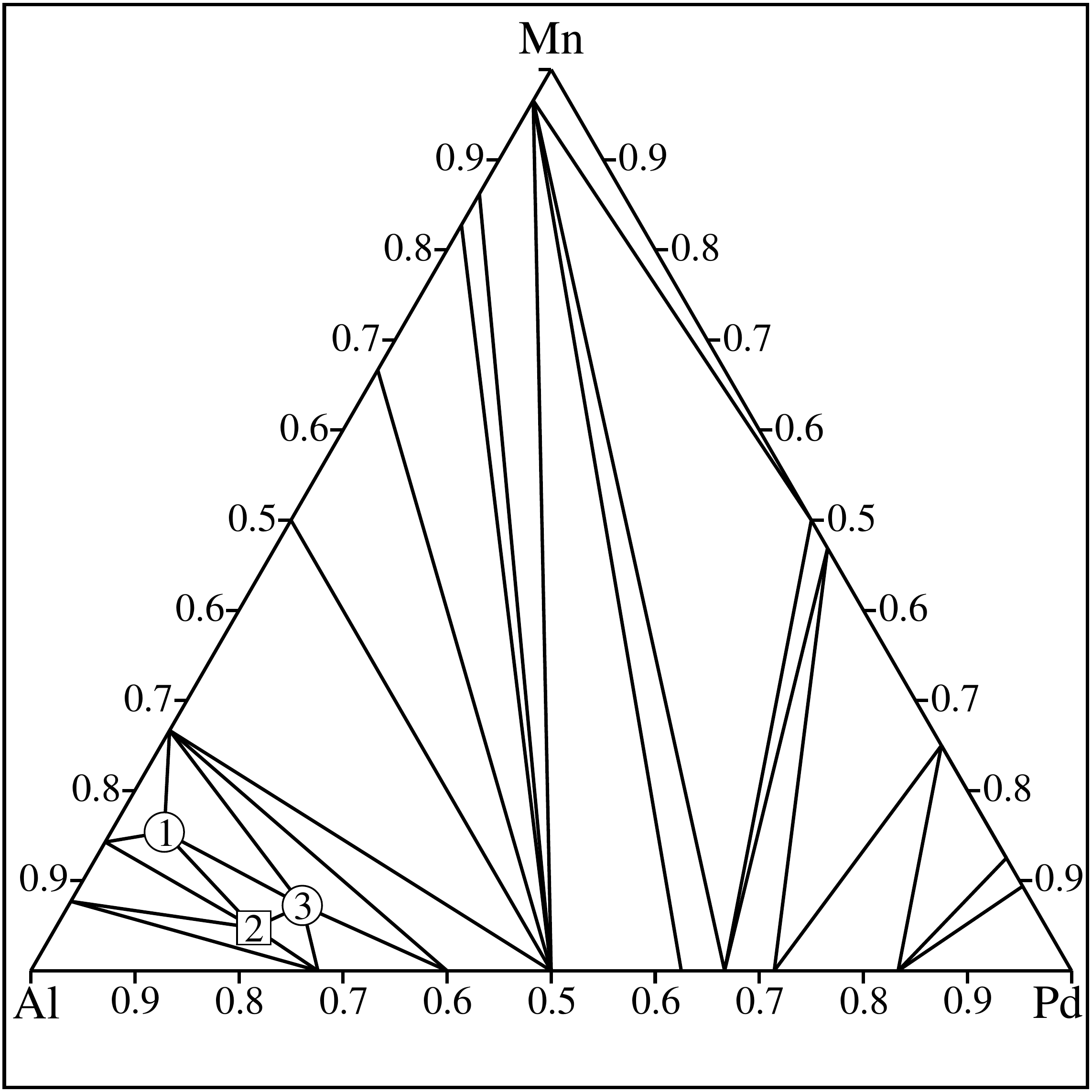}
  \includegraphics[width=0.45\textwidth]{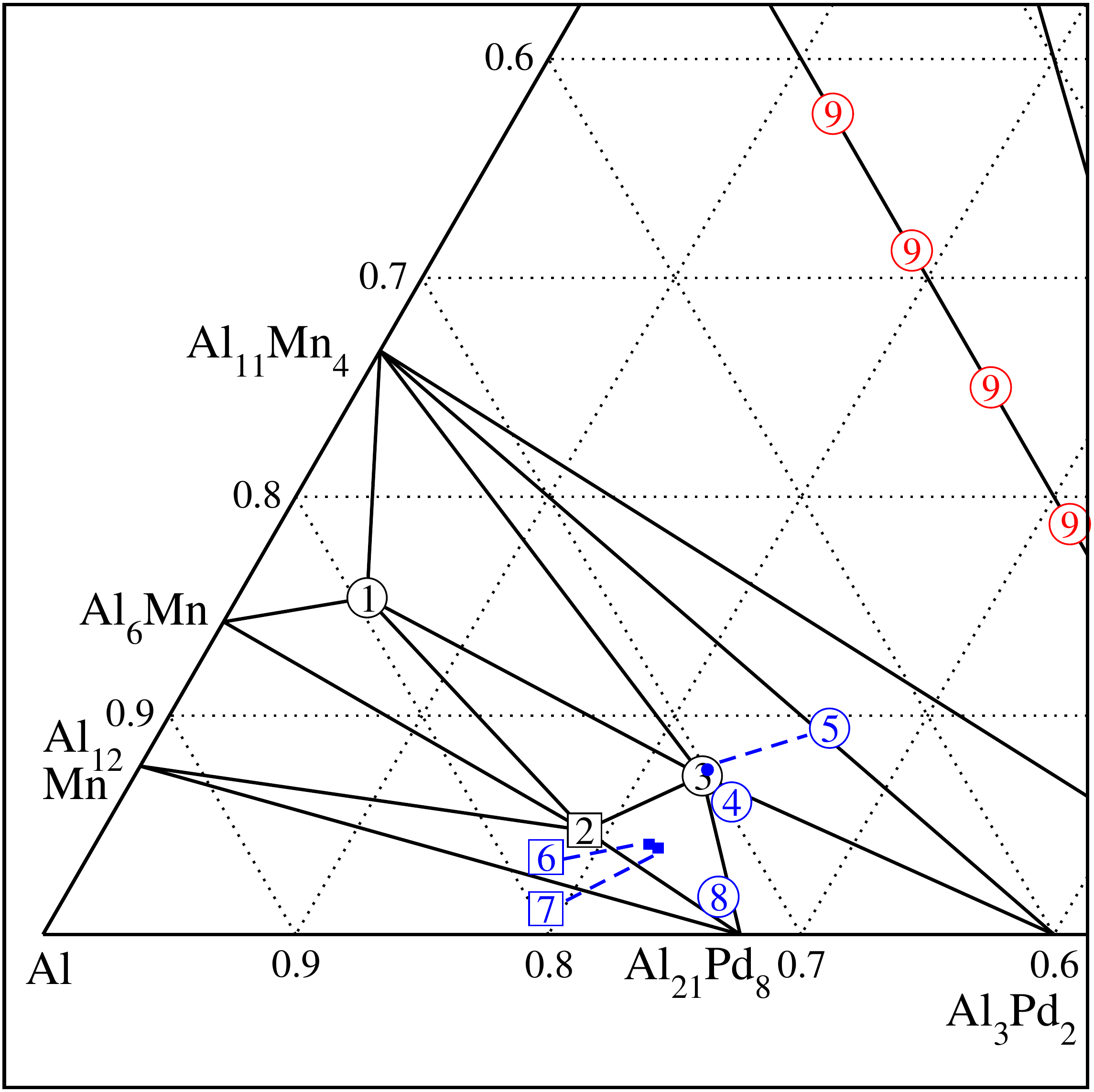}
  \caption{\label{fig:Al-Mn-Pd} Convex hull (a) and Al-rich phases (b) of the Al-Mn-Pd alloy system.  Plotting symbols as in Fig.~\ref{fig:Al-Mn}}
\end{figure}

\begin{table*}[!htbp]
  \caption{\label{tab:Al-Mn-Pd} Selected enthalpies of the Al-Mn-Pd ternary system. Labels specify plotting symbols on Fig.~\ref{fig:Al-Mn-Pd}. ST indicates stable at 840C. Other details as in Table~\ref{tab:Al-Mn}. }
  \begin{tabular}{l|rrr|r|rrrr}
    Label & Phase   & Pearson & Space & ASM       & \multicolumn{3}{c}{DFT} \\
          & name    & symbol  & group & stability & x$_{\rm Mn}$ & x$_{\rm Pd}$ & $\Delta H$ & $\Delta E$ \\
   \hline
   1  & T-Al$_3$(MnPd)   & oP156 & Pnma    & ST & 15.4 &  5.1 & -285 & -11.5 \\
   2  & Al-Mn-Pd-Si     & oP168 & Pnma      & NA &  4.8 & 19.0 & -438 & -2.5 \\
   4  & $i$-1/1 2x2x2   & oP240  & Pbcm     & NA  & 6.7 & 20.0 & -463 & 10.5 \\ 
   4  & $i$-2/1      & aP128  & F$\bar{5}$32 & NA &  7.0 & 21.1 & -493 & 3.3 \\
   3  & $i$-3/2      & aP552  & F$\bar{5}$32 & ST &  7.2 & 22.5 & -523 & -2.7 \\
   5  & $i$-5/3      & aP2338 & F$\bar{5}$32 & NA &  7.5 & 22.5 & -521 & 3.9 \\
   6  & $\epsilon_{16}$ & aP399   & P10$_5$/mmc & NA &  4.0 & 22.1 & -483 & 2.4 \\
   7  & $\xi'$        & oP308    & Pnma     & NA &  3.9 & 22.4 & -489 & 1.8 \\  
   8  & Al$_{21}$Pd$_8$ & tI116 & I4$_1$/a   & ST &  1.7 & 25.9 & -531 & 1.9 \\
   9  & Al$_2$MnPd      & cP2   & Pm$\bar{3}$m & ST & 25.0 & 25.0 & -539 & 41 \\
  \end{tabular}\\
\end{table*}

In contrast to the complexity of the binaries, the reported ternary phase diagram is relatively simple. Its dominant feature is a broad composition range of cubic Al$_2$(Mn,Pd) extending from AlMn up to AlPd. According to our DFT calculations, this phase does not remain stable at low temperatures. Instead, the AlMn.cI2 binary is a disordered high temperature BCC solid solution, and AlPd.cP2 is also a high temperature phase. We also confirm the HT solubility of Mn replacing Pd in the Al$_{21}$Pd$_8$ structure. The convex hull of our calculated enthalpies is shown in Fig.~\ref{fig:Al-Mn-Pd}, along with a close-up of the Al-rich quasicrystal-forming region.

We find that T-Al$_3$(MnPd).oP156, a 12~\AA~ periodic decagonal approximant, is stable in the vicinity of the observed Al-Mn-rich decagonal quasicrystal phase. Indeed, our structure can be recognized as a decagonal quasicrystal approximant. As in the case of its binary counterpart, we find the optimal composition is close to Al$_4$(Mn,Pd). A closely related R-phase (not shown) is nearly degenerate. Electron microscopy images of the decagonal phase reveal $\sim$20~\AA~ decagonal clusters~\cite{hiraga1993} that are too large to be observed in our T- and R-phases, which consist of alternative arrangements of hexagon-shaped tiles.
%Currently, we do not cover a large 12~\AA~ decagonal approximant that would display the characteristic $\sim$20\AA--diameter columnar cluster, clearly emerging in electron microscopy studies~\cite{hiraga1993} -- such study is on the way. In this context, the T or R phase structures only provide atomic motifs decoration for one of the quasicrystal-constituting tiles -- squashed-Hexagon. %% See Fig.1b in hiraga1993; but also MM+Oxborrow model in Aperiodic...

On the Al-Pd-rich side, we find a stable oP168 structure, and coexistence of the icosahedral and 16~\AA-periodic decagonal quasicrystals and their approximants. The stable $oP168$ structure is derived from Sugiyama's quaternary AlMnPdSi.oP168 phase~\cite{Sugiyama} by Si$\rightarrow$Al substitution. At the first glance, this structure looks like a peculiar decagonal relative of the 16~\AA~ decagonal quasicrystal family: viewed down the 24~\AA-periodic $b$ axis, it reveals pentagonal motifs. The $\sim$24~\AA~ period is a 6-fold multiple of the 4~\AA~ periodity characteristic of decagonal Al--TM quasicrystals, and its other lattice parameters relate to the 16~\AA--decagonal approximant $\xi'$ as $c$=7.6~\AA$\sim c_{\xi'}/\tau$ and $a$=14.4$\sim \tau c_{\xi'}$. Here, $\tau$ denotes golden mean (1+$\sqrt{5}$)/2. Closer inspection reveals that the structure contains no icosahedra, but rather Al$_9$Pd, Al$_{10}$Pd and Al$_{10}$Mn clusters, of which none possesses a second shell with a clear pentagonal or icosahedral signature. Thus the most natural relatives of this structure are the binary Al$_4$Pd or Al$_{21}$Pd$_8$ structures.
%% 10.1524/zkri.1998.213.2.90 "Crystal structure of a cubic Al70Pd23Mn6Si; a 2/1 rational approximant of an icosahedral phase" : skipping for now, erroneous refinement

\begin{figure*}
  \includegraphics[width=0.45\textwidth]{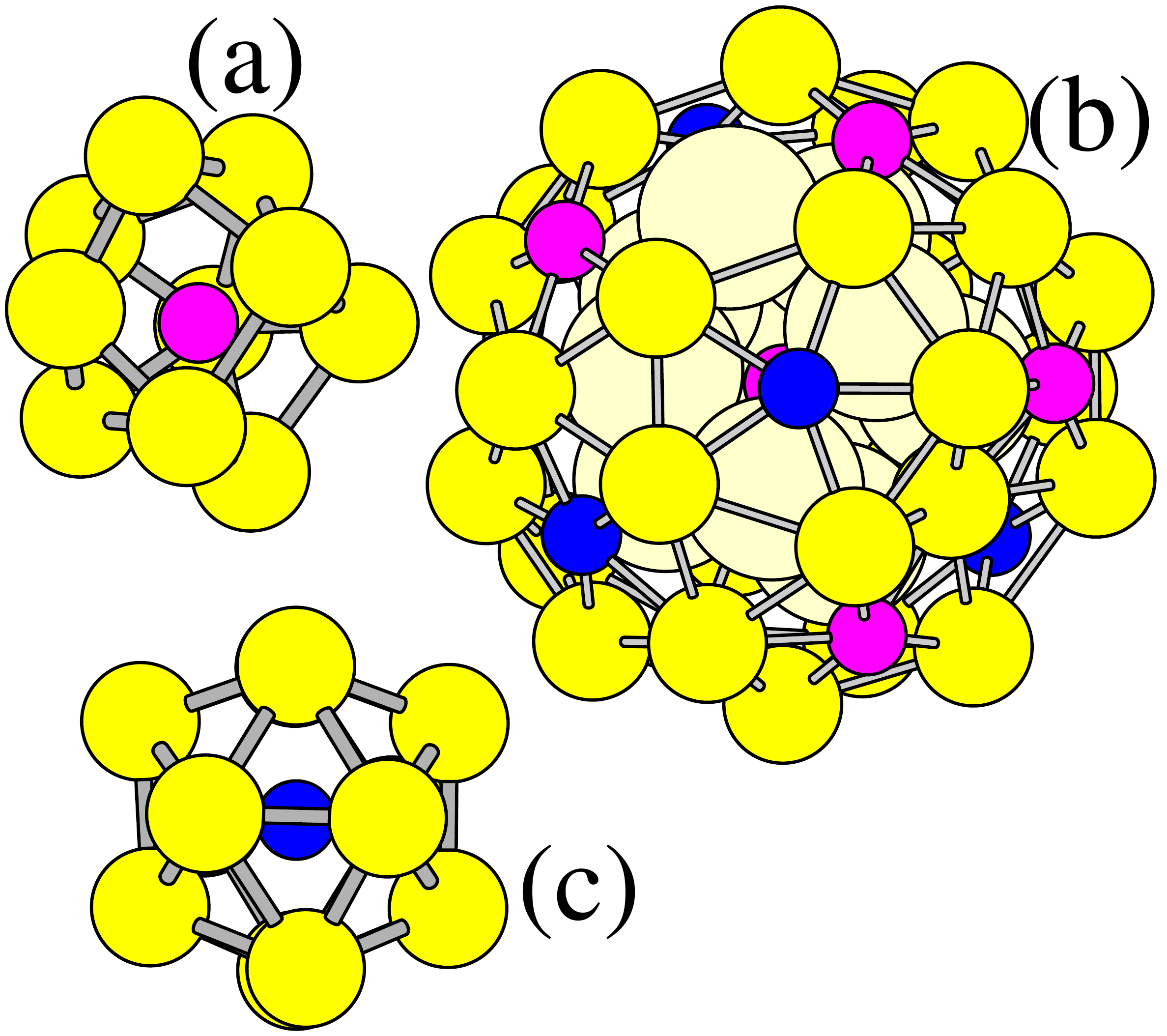}
  \includegraphics[width=0.45\textwidth]{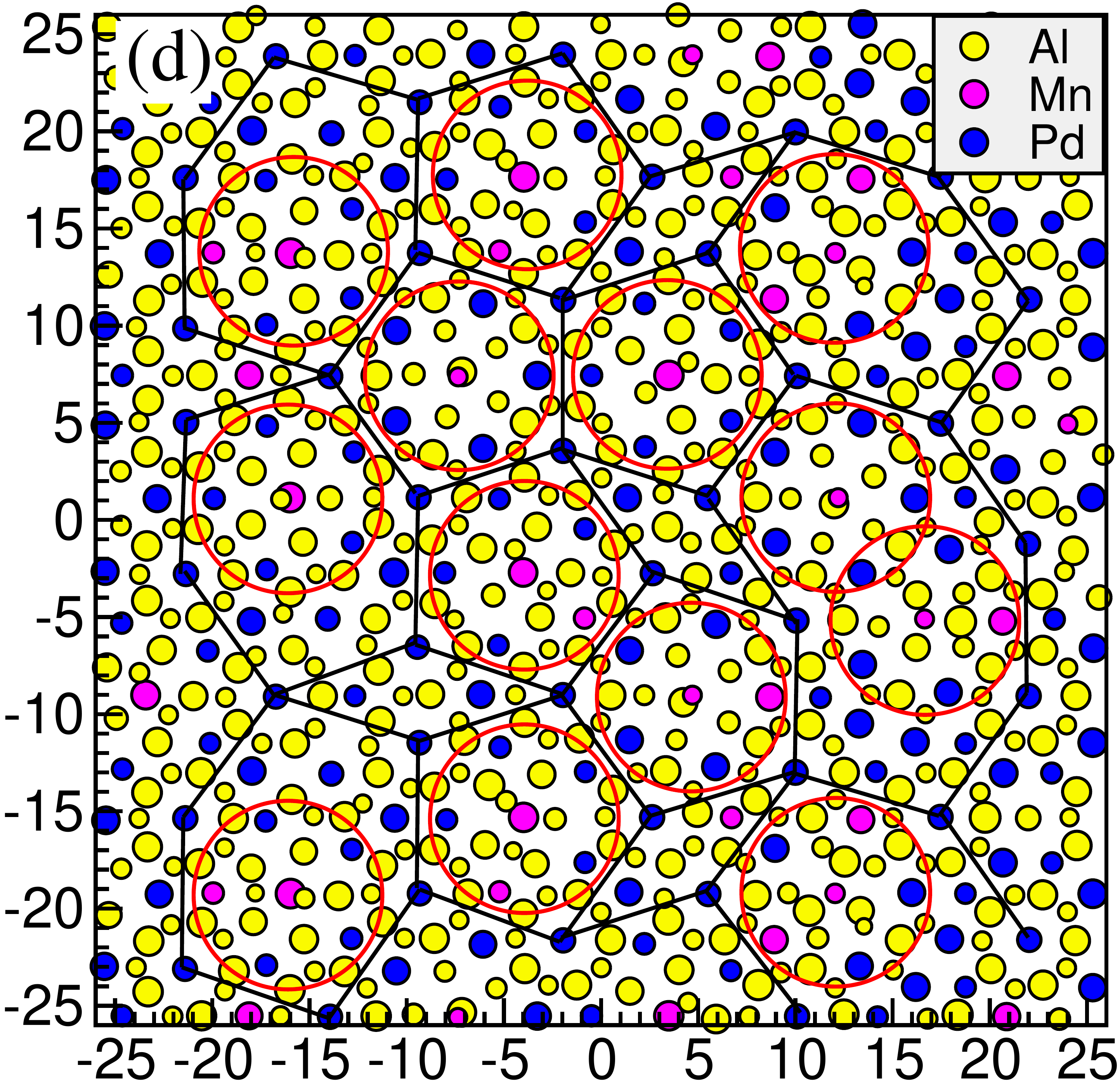}
  \caption{\label{fig:struct} Mn-centered clusters: (a) inner Al$_{9/10}$Mn and (b) outer Mackay shell; (c) Al$_{12}$Pd cluster; (d) 4A thick slice normal to (pseudo) 5-fold axis through 2338-atom "5/3" approximant with  pseudo-Mackay clusters outlined in red circles. A 7.8~\AA~ tiling is shown connecting Al$_{12}$Pd clusters along 2--fold linkages.}
\end{figure*}

``Pseudo-Mackay'' (\pMI) clusters are a fundamental motif of the 16~\AA-decagonal and icosahedral families. They are composed of inner-shell Al$_9$Mn or Al$_{10}$Mn clusters, and an outer shell containing a large (Mn,Pd) icosahedron with an $a_5\sim$4.56~\AA~ 5--fold radius, and 30 Al atoms about half-way between pairs of nearby transition-metal atoms, as in Fig.\ref{fig:struct}a,b. The decagonal family is represented by two nearly stable approximants: $\xi'$ sometimes also denoted $\epsilon_6$, and a larger, ~400 atom/cell approximant denoted $\epsilon_{16}$. The $\xi'$ phase can be grown as a single crystal~\cite{boudard1996}. Its diffraction--refined structure contains unresolved fractional occupancy related to the secondary \pMI structure of the \pMI inner-core shells. The secondary structure was established by $ab-initio$ methods~\cite{Frigan}, and the whole 16~\AA-decagonal family was described as a decorated Hexagon-Boat-Star tiling~\cite{Trebin}. The $\epsilon_{16}$ approximant is the lowest energy large approximant that fairly represents the metastable decagonal quasicrystal phase.

We model icosahedral phase structures in two ways. First, as a decoration of the canonical-cell tiling with 2--fold linkage length $b=2\tau a_5/\sqrt{\tau+2}\sim$7.76~\AA, placing Mn-centred \pMI clusters at even tiling vertices and Al$_{12}$Pd icosahedra ($I$-clusters) at odd vertices. Second, as structures that form spontaneously in our atomistic simulations. We label the approximants with a ratio of successive Fibonacci numbers $F_{n+1}/F_n$,  with 128 atoms for ``2/1'' up to 552 for ``3/2'', and 2338 for ``5/3''. Our naming convention is drawn from the definition of Henley's canonical cells designed for packing icosahedral clusters~\cite{clhcct}. The 2x2x2 supercell of the $i$-1/1 approximant is a decoration of the pure A-cell tiling, inspired by Al$_{11}$Ir$_4$ phase modelling~\cite{MihalkovicHenley}, and it is the only deterministic, regular structure among our icosahedral approximants. The 2/1 approximant is a decoration of the cubic ABC tiling with four \pMI and four $I$-clusters per unit cell, with its secondary structure -- correlations in orientation and variations of the inner-core Al$_{10}$Mn or Al$_9$Mn clusters inside the \pMI -- optimized through simulation at T=500K. The two large approximants 3/2 and 5/3 are chemically disordered structures that spontaneously form at high temperatures in the simulations.

The resulting 3/2 and 5/3 approximant structures prove that the Mn-centred \pMI and $I$ clusters (Fig.\ref{fig:struct}a-c) are the two fundamental building blocks. In the figure, panel (d) shows a spontaneously formed tiling with icosahedra connected along 2--fold linkages; \pMI clusters with Al$_9$Mn or Al$_{10}$Mn inner cores are outlined as red circles. Note that, occasionally, four $I$ clusters form skinny rhombus tile, with short body diagonal distance $b/\tau\sim 4.80$~\AA: such cluster--cluster distance is forbidden in CCT. Also, analysis of the 3/2 and 5/3 approximants reveals that $I$-clusters are more prevalent than \pMI. While the total number of clusters nearly matches the expected number of CCT nodes, the frequency ratio of the two clusters is about $\tau$:1 in favor of Al$_{12}$Pd $I$-clusters, instead of 1:1 as in a CCT--based model. The 2x2x2--3/2 approximant crystal in the quaternary AlPd(CrFe), with around 4400 atoms in cubic cell~\cite{Fujita} -- is entirely tiled by all four kinds of canonical cells, providing a strong argument in favor of the CCT geometry. Thus, while our simulations produce structures with favorable DFT enthalpies, they might not capture more regular (lower entropy, but also lower energy) realizations of the quasicrystal structure.

For comparison, we re-computed total energies of other icosahedral phase models. The Elser's~\cite{QuandtElser} model and Zijlstra's~\cite{Zijlstra} revision of it  (both in 65-atom primitive cells at the same composition) are at +45 and +18 meV/atom respectively. Elser's model shares the same building blocks and 2--fold linkages as CCT, but arranges them on alternating Penrose tiling nodes -- hence \pMI and $I$-clusters are linked by 5--fold $\tau a_{qc}\sim$7.38~\AA~long \pMI-$I$ linkages. The studied approximant is not sufficiently representative, and there remain unprobed aspects of the model that would only appear in larger approximants.

Krajci et al.~\cite{Krajci} applied chemical ordering, inspired by Boudard et al.~\cite{Boudard1992}, to the 6D Katz-Gratias (KG) atomic structure~\cite{Gratias}. The 544-atom ``3/2 approximant'' implementation of the model structure lies at +30 meV/atom after relaxation (not shown). The CCT and 6D model share an underlying cluster network: \pMI--\pMI and $I$--$I$ connect along 2--fold $b$-linkages, and \pMI--$I$ connect along 3--fold $b\sqrt{3}/2$ linkages. Larger approximants possess short 5--fold \pMI-$I$ linkages with length $a_q$. The KG model contains low-coordination Al$_7$Mn local environments instead of Al$_8$Mn or Al$_9$Mn mini-clusters, half of which are inner-cores of the \pMI. Molecular dynamics annealling at moderate temperatures lowers the energy by up to 10 meV/atom, while introducing displacements up to 1.6~\AA. Finally, we note that the $I$ clusters of the KG model form inner shells of the so called ``Bergman'' or ``mini-Bergman'' clusters. The outer dodecahedral shell of this cluster displays chemical symmetry breaking dictated by sharing atoms with nearby \pMI clusters.

\begin{figure*}
  \includegraphics[width=0.45\textwidth]{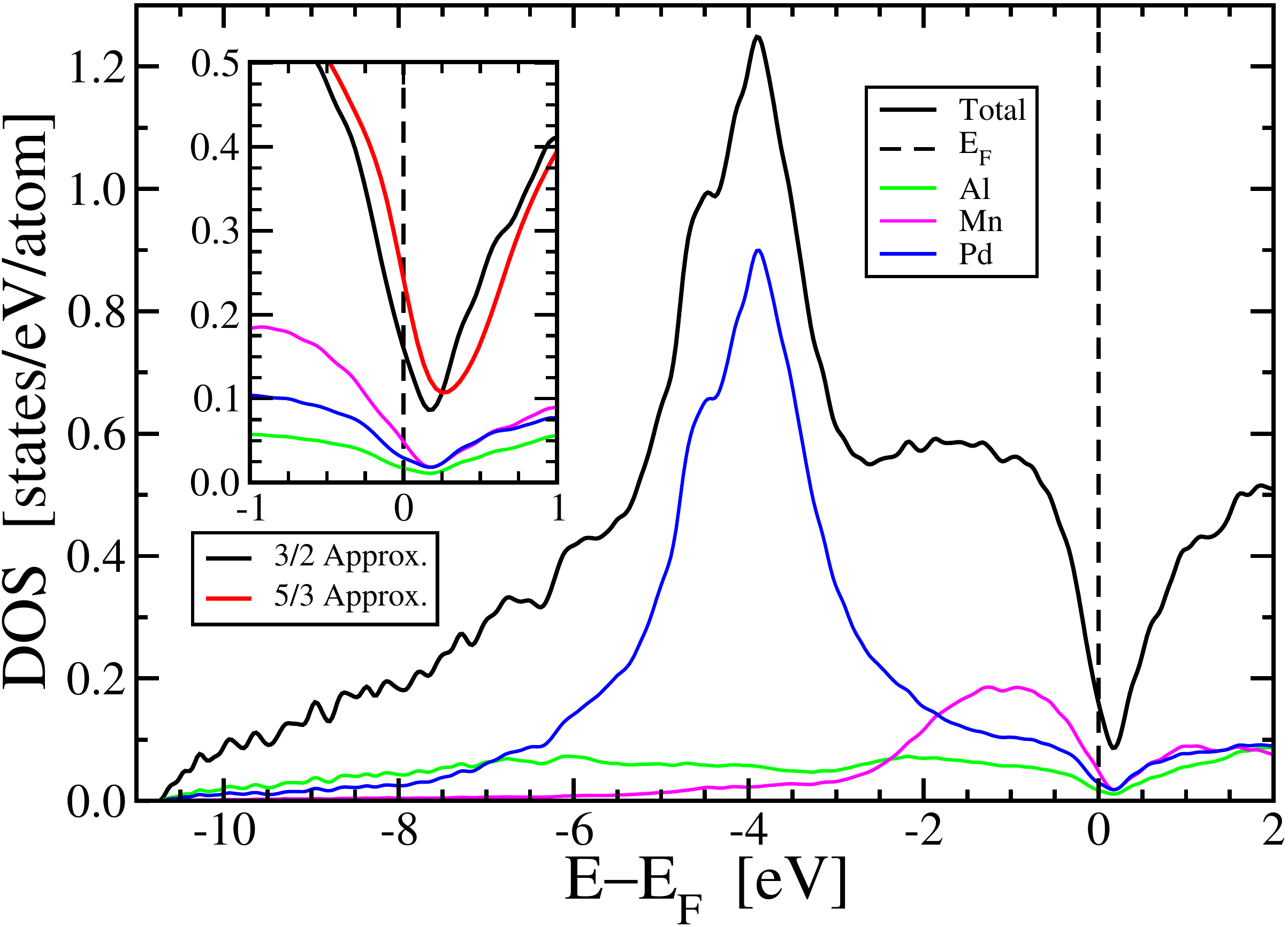}
  \caption{\label{fig:dos} Electronic density of states for Al$_{388}$Mn$_{40}$Pd$_{124}$ (black) and  Al$_{390}$Mn$_{40}$Pd$_{122}$ (inset, red). Partial densities of states show the projected contributions of Al (green), Mn (magenta), and Pd (blue) to the total density of states/atom.}
\end{figure*}

Our stable icosahedral 3/2 approximant has composition Al$_{388}$Mn$_{40}$Pd$_{124}$, for a total of 552 atoms, making it the largest known quasicrystal approximant that is predicted to be LT stable according to DFT. As is frequently observed in Al-rich quasicrystals, the Fermi energy lies within a pseudogap region of the electronic density of states. However, in contrast to the case of i-AlCuFe~\cite{i-AlCuFe}, where $E_F$ lies very close to the pseudogap minimum, in i-AlMnPd $E_F$ lies slightly below minimum in the energetically optimal structure (See Fig.~\ref{fig:dos}). In the 5/3 approximant, with slightly higher energy, the pseudogap is less deep and further above $E_F$, suggesting that it might be possible to further reduce the energy through chemical substitution.

\section{Conclusion}

We have surveyed the stability of most reported structures in the Al-Mn-Pd alloy system and its three binary subsystems. Although we find agreement between calculated formation enthalpies and experimentally assessed phase diagrams in the majority of cases, we find conflicts in roughly 1/3 of binary phases. In the case of the ternary, no reported crystal structure is predicted to be stable, while large quasicrustalline approximants appear to constitute the ternary ground states. Conflicts between experiment and calculation can indicate a failure of the experiments to distinguish low- and high-temperature stability (see, e.g. Ref.~\cite{Wolverton2001}), and metastability.

Alternatively, the conflicts could indicate that the calculated structure incorrect, possibly as the result of incomplete optimization of mixed or partial site occupation or lack of a suitable crystallographic refinement.  Some conflicts may be due to the lack of quantum zero-point and finite temperature corrections. Most of the discreppancies lie beyond the likely calculational errors due to the density functional and pseudopotential approximations.

This work highlights numerous areas requiring further experimental and theoretical investigation. Matching the ASM-assessed phase diagrams with ICSD-reported crystal structures is itself a challenge, but it allows the claims of phase stability to be tested through first principles calculation. This process provides a rich source of important experimental and theoretical research opportunities. Such an approach could be beneficially applied broadly across all alloy systems.

Icosahedral approximant structures may be found online at Ref.~\cite{alloy}.

\section{Acknowledgements}
MM is thankful for the support from the Slovak Grant Agency  VEGA (No. 2/0144/21) and APVV (No. 20-0124, No. 19-0369). MW was supported by the Department of Energy under Grant No. DE-SC0014506. This research also used the resources of the National Energy Research Scientific Computing Center (NERSC), a US Department of Energy Office of Science User Facility operated under contract number DE-AC02-05CH11231 using NERSC award BES-ERCAP24744.

\bibliography{refs}
 
\end{document}